\documentclass[twocolumn,showpacs,prb]{revtex4-1}
\usepackage{graphicx}
\usepackage{dcolumn}
\usepackage{amsmath}
\usepackage{amssymb}

\makeatletter
\def\btt#1{\texttt{\@backslashchar#1}}
\DeclareRobustCommand\bblash{\btt{\@backslashchar}} \makeatother

\begin{document}

\title{Bulk-boundary correspondence in Josephson Junctions}

\author{Jeongmin Yoo$^1$, Tetsuro Habe$^1$, and Yasuhiro Asano$^{1,2}$ }
\affiliation{
$^1$Department of Applied Physics, Hokkaido University, Sapporo 060-8628, Japan\\
$^2$Center for Topological Science \& Technology, Hokkaido University, Sapporo 060-8628, Japan
}

\date{\today}

\begin{abstract}
We discuss Andreev bound states appearing at the interface between two different 
superconductors characterized by different nontrivial topological numbers such as 
one-dimensional winding numbers and Chern numbers. 
The one-dimensional winding number characterizes $d_{xy}$ and $p_x$ wave superconductors. 
The Chern number characterizes chiral superconductors.
The number of interfacial bound states at the zero-energy is equal to the difference between 
the topological numbers on either sides of the Josephson junction.
We discuss also relations between properties of the Andreev bound states at the zero-energy
and features of Josephson current at low temperature.
\end{abstract}


\maketitle

\section{introduction}
According to the topological classification of matter~\cite{kane}, 
a number of unconventional 
superconductors have been categorized in terms of 
nontrivial topological numbers~\cite{schnyder,qi} 
such as $Z_2$ number, Chern number, and one-dimensional winding number. 
The non-centrosymmetric superconductor~\cite{bauer,frigeri} is topologically 
nontrivial when the amplitude of spin-triplet  helical-$p$ wave order parameter is larger than 
that of spin-singlet $s$ wave component~\cite{sato1}. 
Such superconducting phase is characterized by a topological number $Z_2=1$.
The transport properties of non-centrosymmetric superconductors are qualitatively 
different depending on $Z_2$ number~\cite{ncs}.
 The spin-triplet chiral-$p$ wave superconductivity in 
Sr$_2$RuO$_4$~\cite{maeno,rice} is characterized by Chern numbers $n=\pm 1$~\cite{volovik}. 
The Chern numbers here is referred to as Thouless-Kohmoto-Nightingale-den Nijs (TKNN) 
number in solid state physics\cite{tknn}. 
The spin-singlet chiral-$d$ wave ($n=\pm 2$) 
superconductivity has been suggested in 
Na$_x$CoO$_2\cdot y$ H$_2$O~\cite{takeda,baskaran,ogata,atanaka}, heavy fermionic 
compounds~\cite{kasahara,kusunose}, graphene~\cite{nandkishore}, and high-$T_c$ superconductors~\cite{laughlin,belyavsky}.
Unconventional $d_{xy}$ wave symmetry in high-$T_c$ superconductors and 
$p_x$ wave symmetry in the polar state in $^3$He are characterized by the 
one-dimensional winding number~\cite{sato0,sato2} which we call Sato number in this paper. 

The unconventional superconductors have subgap Andreev bound states (ABS) at their 
surface~\cite{buchholtz,hara,hu,tanaka95,ya04}, which has been known for some time.
Such surface state is responsible for 
unusual low energy transport in high-$T_c$ superconductors~\cite{tanaka96,golubov,fogelstrom,lofwander,barash1,hilgenkamp,tafuri},
chiral-$p$ wave superconductor~\cite{barash,laube,ya02,yakovenko,ssro,brydon}.
In particular in spin-triplet superconductors, the surface states attract much 
attention these days because they are recognized as Majorana fermion bound states~\cite{Read,Bolech,Sengupta,Kitaev,ivanov,dassarma}. 
The proximity effect of spin-triplet superconductors is know to be anomalous
because of the penetration of the Majorana bound state into a normal metal~\cite{triplet_odd,ya12}.
 
Today the presence of such surface bound state is explained 
in terms of the bulk-boundary correspondence of topological superconductivity.
According to the bulk-boundary correspondence, 
the number of the surface bound state at the zero-energy would be 
identical to the absolute value of topological number defined in the bulk superconductor. 
In fact, this prediction has been confirmed in a number of theoretical studies.  
In Josephson junctions, the validity of the bulk-boundary correspondence is not clear, 
when the two superconductors are characterized by Sato numbers.

In this paper, we discuss the number of zero-energy ABS at the interface between two 
superconductors belong to different topological classification 
by solving the Bogoliubov-de Gennes equation analytically. 
We first study the interfacial states between two superconductors belonging to different 
Sato numbers. 
Since definition of the Sato number requires the presence of the time-reversal symmetry
(TRS) of the junction, the zero-energy ABS appears only when the phase difference across 
the junction ($\varphi$) is 0 or $\pi$. At $\varphi=0$ or $\pi$,
we confirm that the number of the zero-energy ABS's is equal to the difference of 
Sato numbers in the two superconductors consistently with the bulk-boundary correspondence 
of the topological superconductors. 
We also show that the Josephson current at the zero temperature has large values near 
$\varphi=0$ or $\pi$ because of the resonant tunneling through ABS at the zero-energy. 
Next we confirmed that the number of zero-energy ABS's appearing at the interface 
between two different chiral superconductors is equal to the difference in the
TKNN numbers in the two superconductors. In contrast to $d_{xy}$ and $p_x$ cases, 
the ABS's at the zero-energy do not affect the Josephson current between two chiral 
superconductors at low temperature. We also discuss the stability of $\pi$ state at 
the Josephson junctions just below superconducting transition temperature $T_c$.

This paper is organized as follows. In Sec.~II, we discuss a theoretical model
of Josephson junction consisting two topological superconductors. 
In Sec.~III, we study the interfacial ABS between two superconductors characterized by 
different Sato numbers. The number of the zero-energy ABS and the Josephson effect 
are studied for two chiral superconductors in Sec.~IV.
 We summarize this paper in Sec.~V.

\section{model}
\begin{figure}[tbh]
\begin{center}
\includegraphics[width=8cm]{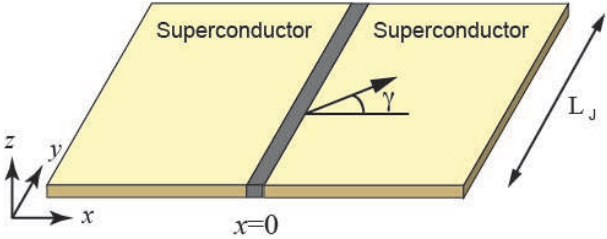}
\end{center}
\caption{(color online). A schematic picture of the Josephson junction. 
}
\label{fig1}
\end{figure}

Let us consider a Josephson junction consisting of two superconductors as shown in Fig.~\ref{fig1}, 
where the electric current flows in the $x$ direction and the junction width in the 
$y$ direction is $L_J$. We apply the periodic boundary condition in the $y$ direction
and consider the limit of $L_J\to \infty$.
The Bogoliubov-de Gennes (BdG) Hamiltonian in momentum space reads
\begin{align}
H_{\textrm{BdG}}(\boldsymbol{k})=&\left[ \begin{array}{cc}
\hat{h}(\boldsymbol{k}) & \hat{\Delta}(\boldsymbol{k}) \\
-\hat{\Delta}^\ast(-\boldsymbol{k}) &-\hat{h}^\ast(-\boldsymbol{k})
\end{array}\right], \label{bdg}\\
\hat{h}(\boldsymbol{k})=& \xi_{\boldsymbol{k}} \hat{\sigma}_0,\quad
\xi_{\boldsymbol{k}}=\frac{\hbar^2\boldsymbol{k}^2}{2m} -\mu,
\end{align}
where $\hat{\sigma}_j$ for $j=1-3$ are the Pauli matrices, $\hat{\sigma}_0$
is the unit matrix in spin space, 
and $\mu$ is the chemical potential.
In this paper, we consider following pair potentials $\hat{\Delta}(\gamma)$,
\begin{align}
\begin{array}{ll}
\Delta i\hat{\sigma}_2 & \qquad \textrm{singlet}\; s,\\
\Delta 2\cos(\gamma)\sin(\gamma)i\hat{\sigma}_2 &\qquad \textrm{singlet}\; d_{xy},\\
\Delta\cos(\gamma)\hat{\sigma}_1 & \qquad\textrm{triplet}\; p_{x},\\
\Delta e^{i n\gamma}i\hat{\sigma}_2 &\qquad\textrm{singlet chiral},\\
\Delta e^{i n\gamma}\hat{\sigma}_1 &\qquad \textrm{triplet chiral},
\end{array} 
\label{sym}
\end{align}
where $\Delta$ is the amplitude of the pair potential, 
$-\pi/2\leq \gamma \leq \pi/2$ is the moving 
angle of a quasiparticle as shown in Fig.~\ref{fig1},
$k_x=k_F\cos\gamma (k_y=k_F\sin\gamma)$ 
is the wavenumber on the fermi surface in the $x$ ($y$) direction, 
and $k_F$ is the Fermi wave number.
\begin{table}[th]
\caption{\label{table1}%
The correspondence between pairing symmetry and the Sato number $W$.
The Sato number can be defined only in the presence of the time-reversal symmetry
for each direction of wave vector on the Fermi surface $\gamma$ and for each Nambu
space. 
The Sato number of $s$ wave case is always zero (i.e, $W_s=0$) because 
$s$-wave superconductor is topologically trivial.  
For spin-singlet $d_{xy}$ symmetry, the Sato number $W_{d_{xy}}$ depends also 
on the Nambu space indicated by $\mathcal{N}1$ and $\mathcal{N}2$.
The Sato number for spin-triplet $p_{x}$ wave case $W_{p_{x}}$ is always unity for all $\gamma$ 
and the two Nambu space.
Here the superconducting phase is take to be zero.
}
\begin{ruledtabular}
\begin{tabular}{lcccc}
\null & angle & $W_s$ & $W_{d_{xy}}$ & $W_{p_x}$ \\
\colrule
$\mathcal{N}1$ & $0< \gamma <\pi/2$ & 0 & 1 & 1 \\
\null & $-\pi/2 < \gamma < 0$ & 0 & $-1$ & 1 \\
\colrule
$\mathcal{N}2$ & $0< \gamma <\pi/2$ & 0 & $-1$ & 1 \\
\null & $-\pi/2 < \gamma < 0$ & 0 & 1 & 1 \\
\end{tabular}
\end{ruledtabular}
\end{table}
The Sato number is defined for each moving angle of a quasiparticle $\gamma$ 
and each Nambu space 
in the presence of TRS.
For spin-singlet superconductors, the BdG Hamiltonian in Eq.~(\ref{bdg}) is block diagonal 
in two Nambu space: $\mathcal{N}1$ and $\mathcal{N}2$. 
In $\mathcal{N}1$, spin of electron-like (hole-like) quasiparticle is $\uparrow$ ($\downarrow$).
On the other hand in $\mathcal{N}2$, spin of electron-like (hole-like) quasiparticle is $\downarrow$ ($\uparrow$).
In this paper, we assume that $\boldsymbol{d}$ vector in the spin-triplet symmetry aligns along
the third axis in spin space. Under this choice, the BdG Hamiltonian in Eq.~(\ref{bdg}) in the spin-triplet cases
is also decoupled into $\mathcal{N}1$ and $\mathcal{N}2$.
In Table.~I, we summarized the Sato number $W(\gamma)$ for $d_{xy}$ and $p_x$ superconductors.

In the chiral states, $n$ in Eq.~(\ref{sym}) must be even integer numbers for 
spin-singlet symmetry, whereas 
it should be odd integers for spin-triplet symmetry. 
The chiral-$p$, $d$ and $f$ wave superconductors
are characterized by the TKNN number $n=\pm 1$, $\pm 2$ and $\pm 3$, respectively.
The TKNN number is defined in the absence of TRS. We note that
the $s$ wave superconductor is topologically trivial. Thus both the Sato number and the TKNN one 
are always zero in the $s$ wave superconductor.

\begin{widetext}
The energy eigen values of Eq.~(\ref{bdg}) are $E=\pm E_{\boldsymbol{k},\pm}$ with 
$E_{\boldsymbol{k},\pm} =\sqrt{ \xi^2_{\boldsymbol{k}} + |\Delta_\pm|^2}$, $\Delta_+=\Delta(\gamma)$,
and $\Delta_-=\Delta(\pi-\gamma)$. All the pair potentials in Eq.~(\ref{sym}) satisfy
$|\Delta_+|=|\Delta_-|$. 
In such case, the wave functions in the left and the right superconductors in $\mathcal{N}1$
are obtained as~\cite{ya01}
\begin{align}
\Psi_L(x,y) =&  \hat{\Phi}_L
\left[ 
\left[
\begin{array}{c}
u_{L} \\
v_{L}s_{L+}^\ast 
\end{array} 
\right]
a e^{ik_L^{e}x}
+\left[
\begin{array}{c} 
{v}_{L} s_{L-}\\ 
{u}_{L} 
\end{array} \right]
b e^{-ik_L^{h}x}
+
\left[
\begin{array}{c}
u_{L} \\
v_{L}s_{L-}^\ast 
\end{array} 
\right]
A e^{-ik_L^{e}x}
+
\left[
\begin{array}{c} 
{v}_{L}s_{L+}\\ 
{u}_{L} 
\end{array} \right]
B e^{ik_L^{h}x}
\right] 
 e^{ik_yy},\label{phil}\\
\Psi_R(x,y) =&  \hat{\Phi}_R
\left[
\left[
\begin{array}{c}
u_{R} \\
v_{R}s_{R+}^\ast 
\end{array} 
\right]
C e^{ik_R^{e}x}
+\left[
\begin{array}{c} 
{v}_{R}s_{R-}\\ 
{u}_{R}
 \end{array} 
 \right] D e^{-ik_R^{h}x}
\right] 
 e^{ik_yy},\label{phir}
\end{align}
\begin{align}
u_{j} =&\sqrt{ \frac{1}{2}\left(1+\frac{\Omega_{j}}{E}\right)}, \; 
v_{j} =\sqrt{\frac{1}{2}\left(1-\frac{\Omega_{j}}{E}\right)}, \;
\Omega_{j}= \sqrt{ E^2-|\Delta_{j}|^2}, \; 
s_{j\pm}= \frac{ \Delta_{j\pm}}{|\Delta_{j\pm|}},\; 
 \hat{\Phi}_{j}=\text{diag}\left\{ e^{i\varphi_j/2},
e^{-i\varphi_j/2} \right\}, \\
k_j^{e}=& \left[ k_x^2 + \frac{2m}{\hbar^2}\sqrt{E^2-|\Delta_{j}|^2} \right]^{1/2},\;
k_j^{h}= \left[ k_x^2 - \frac{2m}{\hbar^2}\sqrt{E^2-|\Delta_{j}|^2} \right]^{1/2},
\end{align}
where $j=L$ ($R$) indicates the left (right) superconductor and 
 $\varphi_j$ is the macroscopic phase of the superconductor. 
The coefficients $A$, $B$, $C$, and $D$ are the amplitudes of outgoing waves from the 
interface and $a$ and $b$ are those of incoming waves.
At the junction interface, we introduce the potential barrier 
described by $V_0\delta(x)$. 
The boundary conditions for wave function become
\begin{align}
\Psi_L(0,y) =& \Psi_R(0,y), \quad -\frac{\hbar^2}{2m}\left[ \left.\frac{d}{dx}\Psi_R(x,y)\right|_{x\to 0^+}
-\left.\frac{d}{dx}\Psi_L(x,y)\right|_{x\to 0^-} \right]
+V_0 \Psi_R(0,y)=0.\label{bc}
\end{align}

When we calculate the energy of the interfacial ABS, we put $a=b=0$.
Since we seek the ABS's for $|E| < |\Delta_{j}|$, 
$\Psi_L(x,y)$ ($\Psi_R(x,y)$) decays at $x\to -\infty (\infty)$.
The decay length is given by the coherence length $\xi_0\approx \hbar v_F/ (\pi \Delta)$ 
with $v_F$ being the fermi velocity. 
By using the boundary conditions in Eq.~(\ref{bc}), we obtain the relation among
$A$, $B$, $C$, and $D$ as
$\check{\mathcal{Y}}[A,B,C,D]^{t}=0$, where $[\cdots]^{t}$ is the transpose of $[\cdots]$
and $\check{\mathcal{Y}}$ is a $4 \times 4$ matrix calculated from the boundary conditions.
We define that $\Xi(E)$ is the determinant of $\check{\mathcal{Y}}$. It is expressed by 
\begin{align}
\Xi(E) \equiv &(u_{L}^2-v_{L}^2 s_{L+} s_{L-}^\ast)
(u_{R}^2-v_{R}^2 s_{R+}^\ast s_{R-}) \nonumber\\
+&|t_n|^2 
\left\{
u_{L}^2v_{R}^2 s_{R+}^\ast s_{R-} + v_{L}^2 s_{L+}s_{L-}^\ast u_{R}^2
- \nu e^{i\varphi} u_{L}v_L s_{L+} u_{R}v_{R} s_{R+}^\ast
- \nu e^{-i\varphi} u_{L}v_{L} s_{L-}^\ast u_{R} v_{R} s_{R-} \right\},\label{eq_abs}
\end{align}
where $\varphi=\varphi_L-\varphi_R$ ($-\pi< \varphi \leq \pi$) is the 
phase difference across the junction
and $t_n = \cos\gamma /( \cos\gamma + iz_0) $ is the normal transmission coefficient
of junction with $z_0= V_0/\hbar v_F$. 
The energy of the ABS's is calculated from a condition $\Xi(E)=0$.

When we discuss the Josephson current, we first calculate the reflection coefficients
of the junction. By eliminating $C$ and $D$
using the boundary conditions in Eq.~(\ref{bc}), we obtain a relation between 
$(a,b)$ and $(A,B)$
as
\begin{align}
\left( \begin{array}{c} A \\B \end{array} \right)
=\left( \begin{array}{cc} 
r_{ee} & r_{eh} \\ 
r_{he} & r_{hh} 
\end{array} \right)
\left( \begin{array}{c} a\\b \end{array} \right),
\end{align}
where $r_{he}$ and $r_{eh}$ are the Andreev reflection coefficients and 
 $r_{ee}$ and $r_{hh}$ are the normal reflection coefficients.
The Josephson current is calculated based on a formula~\cite{ya01}
\begin{align}
J=&\frac{e}{2\hbar}T\sum_{\omega_n} \sum_{\gamma} \sum_{\nu} \left.
\frac{\nu}{\Omega_L}\left[ 
\Delta_{L+} r_{he}- \Delta_{L-}^\ast r_{eh} \right]\right|_{E\to i\omega_n},\label{josephson}\\
r_{he}=&\frac{1}{\Xi}\left[ |t_n|^2
\left\{
e^{i\varphi} u_L^2 u_R v_R s_{R+}^\ast + e^{-i\varphi}v_L^2 s_{L+}^\ast s_{L-}^\ast u_Rv_R s_{R-}
-\nu u_Lv_L s_{L+}^\ast v_R^2 s_{R+}^\ast s_{R-} - \nu u_L v_L s_{L-}^\ast u_R^2 \right\} \right.\nonumber\\
&\left.
+\nu u_L v_L\left\{ s_{L-}^\ast -s_{L+}^\ast \right\}
\left\{ u_R^2-v_R^2 s_{R+}^\ast s_{R-} \right\} \right],\\
r_{eh}=&\frac{1}{\Xi}\left[ |t_n|^2
\left\{
e^{i\varphi} u_L^2 u_R v_R s_{R-} + e^{-i\varphi}v_L^2 s_{L-} s_{L+} u_Rv_R s_{R+}^\ast
-\nu u_Lv_L s_{L-} v_R^2 s_{R+}^\ast s_{R-} - \nu u_L v_L s_{L+} u_R^2 \right\} \right.\nonumber\\
&\left.
+\nu u_L v_L\left\{ s_{L+} -s_{L-} \right\}
\left\{ u_R^2-v_R^2 s_{R+}^\ast s_{R-} \right\} \right],  \label{reh}
 \end{align}
where $\omega_n=(2n+1)\pi T$ is the Matsubara frequency.
\end{widetext}
Here we explain the definition of $\nu$ appearing in Eqs.~(\ref{eq_abs})-(\ref{reh}). 
When the two superconductors are in the spin-singlet symmetry, $\nu$ become 1 for the two Nambu
space $\mathcal{N}1$ and $\mathcal{N}2$. 
This is also true when the two superconductors are in the spin-triplet symmetry.
In these cases, the energy of ABS obtained from Eq.~(\ref{eq_abs}) is degenerate 
in the two Nambu space. As a result, the $\sum_{\nu}$ gives rise a factor 2 in the Josephson 
current in Eq.~(\ref{josephson}).
When one superconductor is in the spin-singlet symmetry and the other is in the spin-triplet one,
we take $\nu=1$ in $\mathcal{N}1$ and $\nu=-1$ in $\mathcal{N}2$. 

The Josephson current can be decomposed into a series of
\begin{align}
J=\sum_{n=1}^{\infty} J_n \sin(n\varphi) + I_n \cos(n\varphi). \label{deco}
\end{align}
 When two superconductors preserve TRS, $I_n$ is 
usually zero. 
The normal transmission probability is defined by
\begin{align}
T_N = \frac{1}{2}\int^{\pi/2}_{-\pi/2}\!\!\! d\gamma \cos\gamma |t_n|^2.
\end{align}
The coefficients $J_n$ is, roughly speaking, proportional to $(T_N)^n$ for $s$ wave junction.
The lowest coupling $J_1$ is sensitive to the pairing symmetries of two superconductors.
For instance, $J_1=0 $ when one superconductor is spin-singlet and the other 
is spin-triplet.

\section{Sato number}

When the two superconductors are in the $s$ wave symmetry, we obtain the well known 
results of the energy of Andreev bound states $E^{\textrm{ABS}}=\pm \epsilon_{s/s}$ 
and the Josephson current $J_{s/s}$
\begin{align}
\epsilon_{s/s}=& \Delta \sqrt{1-|t_n|^2\sin^2(\varphi/2)},\\
J_{s/s}=&J_0 \frac{\Delta_{ }}{\Delta_0} \frac{1}{2T_N} \int^{\pi/2}_{-\pi/2}\!\!\! d\gamma \, 
{|t_n|^2 \cos\gamma \sin\varphi}\; A(\epsilon_{s/s}),\\ 
A(\epsilon)=& \frac{\Delta}{\epsilon}\tanh \left[ \frac{\epsilon}{2T}\right], \label{def_a}\\
J_0 =& \frac{\pi\Delta_0}{2eR_N},\; \; \frac{1}{R_N}=\frac{2e^2}{h}T_N N_c,\;\; N_c=\frac{Wk_F}{\pi}
\end{align}
where $\Delta_0$ is the amplitude of pair potential at $T=0$ and $N_c$ is the number of 
propagating channels at the fermi level. 
The amplitude of the critical current at $T=0$ is $J_0$ in the $s$ wave 
junctions~\cite{ambegaokar}.
There is no zero-energy ABS in the $s/s$ junctions because the $s$ wave superconductor is always 
topologically trivial. 

\begin{figure}[tbh]
\begin{center}
\includegraphics[width=8cm]{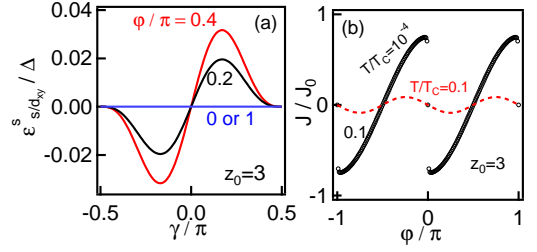}
\end{center}
\caption{(color online). 
The results for $s/d_{xy}$ junctions.
The energy of the Andreev bound state $\epsilon^s_{s/d_{xy}}$ is 
plotted as a function of $\gamma$ at $z_0=3$ in (a).
The current-phase relationship is shown in (b).
}
\label{fig2}
\end{figure}

When $s$ wave superconductor is on the left and $d_{xy}$ wave one is on the right, 
the equation for the energy of the Andreev bound states is obtained as
\begin{align}
2E\sqrt{\Delta^2-E^2}-\Delta^2|t_n|^2|\theta_d|\sin\varphi=0.
\end{align} 
Here we approximately changes $\sqrt{\Delta^2 \theta_d^2 -E^2}$ to $\sqrt{\Delta^2-E^2}$
 with $\theta_d= \sin(2\gamma) $. 
This approximation is possible because the presence or absence of the zero-energy ABS is sensitive 
 only to the sign changing of pair potential on the fermi surface~\cite{ya04}. 
The energy of the ABS in $\mathcal{N}1$ and that in $\mathcal{N}2$ are identical to each other. 
We find that $E^{\textrm{ABS}}=\epsilon_{s/d_{xy}}^s$ and $\epsilon_{s/d_{xy}}^c$ with
\begin{align}
&\epsilon_{s/d_{xy}}^s= {\Delta}
\sin\left( \frac{\alpha_{s/d_{xy}}}{2} \right),\\
&\epsilon_{s/d_{xy}}^c = {\Delta}
\cos\left( \frac{\alpha_{s/d_{xy}}}{2} \right)
\textrm{sgn}\left[\sin\left( \frac{\alpha_{s/d_{xy}}}{2} \right)\right],\\
&\sin(\alpha_{s/d_{xy}}) = |t_n|^2 \theta_d \sin\varphi.
\end{align}
Since $-\pi/2 \leq \alpha_{s/d_{xy}}\leq\pi/2$, $\epsilon_{s/d_{xy}}^c$
does not become zero.
At $|t_n| \to 0$, the two superconductors are separated from each other
and $\epsilon^s_{s/d_{xy}}=0$ for all $\gamma$. 
Such zero-energy ABS corresponds to the surface bound state of $d_{xy}$ wave 
superconductor. 
Namely there is one zero-energy surface state for each Nambu space for each $\gamma$, 
which is a result of the bulk-boundary correspondence of isolated superconductors. 
At $|t_n|\neq 0$, a zero-energy state appears for each $\gamma$ and each Nambu space 
only when $\varphi=0$ and $\pi$. 
In Fig.~\ref{fig2}(a), we show $\epsilon^s_{s/d_{xy}}$ as a function of $\gamma$ 
for several choices of $\varphi$ at $z_0=3$. 
The number of the zero-energy ABS at $\varphi=0$ is equal 
to $|W_s(\gamma)-W_{d_{xy}}(\gamma)|=1$ for each Nambu space.
The Sato number can be defined only in the presence of TRS in the junction under consideration.
The zero-energy ABS disappear for $\varphi\neq 0$ because the Sato number is not be well 
defined in the absence of TRS. 
In fact, the Josephson current flows through the junction interface for $\varphi\neq 0$ as
\begin{align}
J_{s/d_{xy}}=& J_0 \frac{\Delta_{ }}{\Delta_0} 
\frac{1}{2T_N} \int^{\pi/2}_{-\pi/2}\!\!\! d\gamma \frac{\cos\gamma|t_n|^4 \theta_{d}^2 \sin(2\varphi)}{4\cos(\alpha_{s/d_{xy}})}\nonumber\\
&\times\left[ 
A(\epsilon_{s/d_{xy}}^c) - A(\epsilon_{s/d_{xy}}^s)
\right].
\label{j_sdxy}
\end{align}
The current-phase relationship (CPR) for $T \lesssim T_c$ becomes
\begin{align}
J_{s/d_{xy}}=& -  \sin(2\varphi) \nonumber\\
\times& J_0 \frac{\Delta}{\Delta_0}\left(\frac{\Delta}{T} \right)^3 
\frac{1}{2T_N} \int^{\pi/2}_{-\pi/2}\!\!\! d\gamma\frac{\cos\gamma|t_n|^4 \theta_{d}^2}{96}. 
\end{align}
The lowest order coupling is absent (i.e., $J_1=0$) because $s$ and $d_{xy}$ wave pairing function 
is orthogonal to each other.
At $T=0$, the Josephson current becomes
\begin{align}
J_{s/d_{xy}}=& -  \cos(\varphi) \textrm{sgn}[ \sin(\varphi)] \nonumber \\
\times & J_0 \frac{1}{2T_N} \int^{\pi/2}_{-\pi/2}\!\!\! d\gamma 
\frac{\cos\gamma|t_n|^2 |\theta_{d}|}
{\sqrt{1+|t_n|^2 |\theta_{d}\sin(\varphi)|}}. \label{sdxy2}
\end{align}
The resonant transmission through the zero-energy ABS's
 mainly contributes to the Josephson current at $T=0$. 
In fact, the Josephson current becomes large at $\varphi=0$ and $\pi$
and jumps as shown in Fig.~\ref{fig2}(b), 
where we plot the Josephson current as a function of $\varphi$. 
We fix $z_0$ at 3, which leads to the normal transmission probability $T_N\approx 0.07$.
At $T=0$, the amplitude of the Josephson current is roughly proportional to $T_N$ 
as shown in Eq.~(\ref{sdxy2}), which is a result of the resonant transmission 
through the ABS at the zero-energy.

The similar conclusion is obtained when we replace the spin-singlet $d_{xy}$ wave 
superconductor by the spin-triplet $p_{x}$ 
wave one. The energy of ABS is obtained as
$E^{\textrm{ABS}}= \epsilon^s_{s/p_{x}}$ and $\epsilon^c_{s/p_{x}}$.
These energy are given by
\begin{align}
&\epsilon_{s/d_{xy}}^s= {\Delta}
\sin\left( \frac{\alpha_{s/p_{x}}}{2}\right),\\
&\epsilon_{s/d_{xy}}^c = {\Delta}
\cos\left( \frac{\alpha_{s/p_{x}}}{2}\right)
\textrm{sgn}\left[ \sin\left(\frac{\alpha_{s/p_{x}}}{2} \right)\right],\\
&\sin(\alpha_{s/p_{x}}) =  \nu|t_n|^2 \theta_p \sin\varphi,
\end{align}
with $\theta_p=\cos\gamma$.
We note that there are four dispersion branches in the ABS because 
of the contributions from two Nambu space.
The zero-energy ABS appears only when $\varphi=0$ and $\pi$.
The Josephson current in this junction is given by Eq.~(\ref{j_sdxy}) with 
$\theta_d \to \theta_p$.

\begin{figure}[tbh]
\begin{center}
\includegraphics[width=8cm]{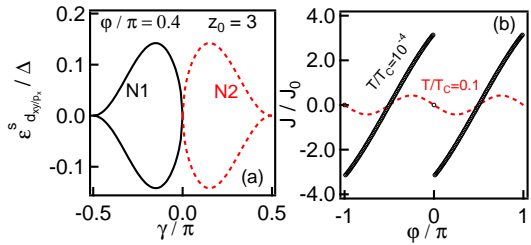}
\end{center}
\caption{(color online). 
The results for $d_{xy}/p_x$ junctions at $z_0=3$.
The energy of the Andreev bound state $\pm \epsilon^s_{d_{xy}/p_x}$ is 
plotted as a function of $\gamma$ in (a).
The current-phase relationship of the Josephson current is shown in (b), where 
results for $T=0.1T_c$ is amplified by 10.
}
\label{fig3}
\end{figure}

\begin{figure}[tbh]
\begin{center}
\includegraphics[width=8cm]{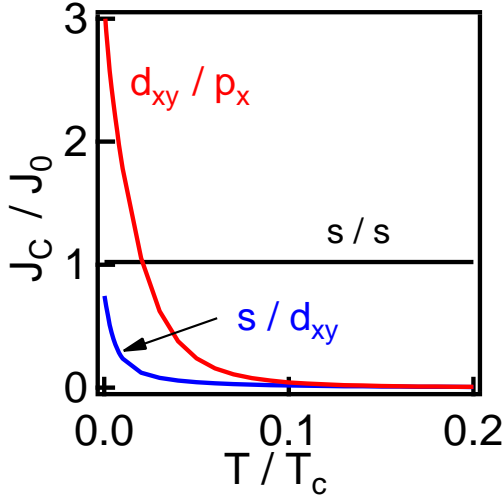}
\end{center}
\caption{(color online). 
The maximum value of the Josephson current is plotted as a function of temperature. 
Here we choose $z_0=3$.
}
\label{fig4}
\end{figure}

As shown in the Table.~\ref{table1}, $d_{xy}$ wave and $p_x$ wave symmetries are classified into the 
same Sato number for $0<\gamma<\pi/2$ in $\mathcal{N}1$ and $-\pi/2<\gamma<0$ in $\mathcal{N}2$.
In these cases, ABS's may not appear at the zero-energy. On the other hand, two zero-energy ABS are expected for 
$0<\gamma<\pi/2$ in $\mathcal{N}2$ and $-\pi/2<\gamma<0$ in $\mathcal{N}1$ because of
 $|W_{d_{xy}}-W_{{p_x}}|=2$. 
These prediction can be confirmed by considering the equation for energy of the ABS,
\begin{align}
E^2-&\frac{|t_n|^2}{2}\left[ E^2+\sqrt{ \Delta^2\theta_d^2 -E^2}
\sqrt{ \Delta^2\theta_p^2 -E^2} \right.\nonumber\\
&\left.+ \nu \Delta^2\theta_p\theta_d \cos(\varphi) \right]=0,
\end{align}
where $\nu= 1$ for $\mathcal{N}1$ and $\nu=-1$ for $\mathcal{N}2$.
By substituting $E=0$, we find $|\theta_d|+\nu \theta_d \cos\varphi=0$.
Therefore, at $\varphi=0$, the zero-energy ABS appear for $0< \gamma < \pi/2$ in $\mathcal{N}2$, 
and for $-\pi/2 < \gamma < 0$ in $\mathcal{N}1$. 
 To have expressions of $E^{\textrm{ABS}}$, we
 apply the approximation $\sqrt{ \Delta^2\theta_d^2 -E^2}
\sqrt{ \Delta^2\theta_p^2 -E^2} \approx \Delta^2|\theta_p\theta_d|-E^2$. 
As a result, we obtain $E^{\textrm{ABS}}= \pm\epsilon_{ d_{xy}/p_x}^{c}$ and
$\pm\epsilon_{ d_{xy}/p_x}^{s}$ with
\begin{align}
\epsilon_{d_{xy}/p_x}^c=&\Delta |t_n| \sqrt{ |\theta_p\theta_d| } \cos(\varphi/2)\Theta(\nu\theta_d),\\
\epsilon_{d_{xy}/p_x}^s=&\Delta |t_n| \sqrt{ |\theta_p\theta_d| } \sin(\varphi/2)\Theta(-\nu\theta_d),
\end{align}
where $\Theta(x)$ is the step function: $\Theta=1$ for $0<x$ and $\Theta=0$ otherwise.
In Fig.~\ref{fig3}(a), we show $\pm \epsilon^{s}_{d_{xy}/p_x}$ at $z_0=3$ and $\varphi=0.4\pi$.
When we consider $\varphi\to 0$, there are doubly degenerate ABS's at the zero-energy 
for $-\pi/2<\gamma<0$ in $\mathcal{N}1$
and for $0<\gamma<\pi/2$ in $\mathcal{N}2$. 
The doubly degenerate dispersionless ABS's at the zero-energy drastically affect 
the Josephson current which is calculated to be 
\begin{align}
J_{d_{xy}/p_x}=&J_0 \frac{\Delta_{ }}{\Delta_0} \sum_{\nu}
\frac{1}{2T_N} \int^{\pi/2}_{-\pi/2}\!\!\! d\gamma 
\frac{\cos\gamma|t_n|^2|\theta_d\theta_p| \sin\varphi}{2} \nonumber\\
&\times
\left[
A(\epsilon_{d_{xy}/p_x}^c) - A(\epsilon_{d_{xy}/p_x}^s) 
\right]. 
\end{align}
The expression of the Josephson current is given by 
\begin{align}
J_{d_{xy}/p_x}=& -\sin(2\varphi) \nonumber\\
\times J_0 \frac{\Delta_{ }}{\Delta_0} 
\left(\frac{\Delta}{T}\right)^3 &\frac{1}{2T_N} \int^{\pi/2}_{-\pi/2}\!\!\! d\gamma
 \frac{\cos\gamma |t_n|^4\theta_d^2\theta_p^2}{96},
\end{align}
for $T \lesssim T_c$. The lowest coupling vanishes because one superconductor is 
spin-singlet and the other is spin-triplet (i.e., $J_1=0$ in Eq.~(\ref{deco})).
At $T=0$, we find 
\begin{align}
J_{d_{xy}/p_x}=&[ \sin(\varphi/2) -\cos(\varphi/2)\textrm{sgn}(\varphi)]\nonumber\\
\times & J_0 \frac{1}{2T_N} \int^{\pi/2}_{-\pi/2}\!\!\! d\gamma
{\cos\gamma|t_n|\sqrt{|\theta_d\theta_p|}},\label{dxypx2}
\end{align}
In Fig.~\ref{fig3}(b), we plot the Josephson current as a function of $\varphi$ 
at $z_0=3$. At very low temperature, the Josephson current shows 
unusual current-phase relationship 
and has large amplitude proportional 
to $\sqrt{T_N}$ around $\varphi=0$.
At $\varphi=\pi$, 
 $\pm \epsilon^{c}_{d_{xy}/p_x}$ describes the 
 doubly degenerate dispersionless ABS at the zero-energy 
which also contribute to the large Josephson current as shown in Fig.~\ref{fig3}(b).
Tuning of $\varphi$ at $\pi$ is equivalent to changes the sign of $W_{d_{xy}}$ in Table.~I.
In Fig.~\ref{fig4}, we show the critical Josephson current as a function of 
temperature for $0\leq T\leq 0.2T_c$ with $z_0=3$.
In this temperature range, the maximum value of $J_{s/s}$ is saturates at $J_0$.
The results for $J_{s/d_{xy}}$ and $J_{d_{xy}/p_x}$ are much smaller than $J_0$ 
for $T>0.1T_c$. They increases rapidly with decreasing temperature for $T<0.1T_c$.
Such effect is known as the low-temperature anomaly of the Josephson 
current~\cite{tanaka96,barash1}. 

\section{TKNN number}
In this section, we consider a junction which consists of two chiral superconductors,
where the TKNN number in the left superconductor is $n$ and that in the right one is $m$.
We can calculate the energy of ABS and the Josephson current without 
any further 
approximations because the pair potential is independent of $\gamma$ and TKNN numbers.
We first show the results of the energy of ABS and the 
Josephson current.
Secondly we count the number of zero-energy ABS of the junction. 
Then we discuss properties of the Josephson current.

We show the results for a junction which consists of two spin-singlet chiral 
superconductors. 
In this case, the two TKNN numbers ($n$ in the left superconductors and $m$ in the right one) 
are even integer numbers.
The energy of ABS's are obtained as
$E^{\textrm{ABS}}=\epsilon^{ee}_{\pm}$ with,
\begin{align}
\epsilon^{ee}_\pm =&\pm \Delta\textrm{sgn}[ \sin(X^{ee}_\pm ) ] \cos(X^{ee}_\pm),\label{ee}\\
X^{ee}_{\pm}=& \frac{\alpha^{ee}\pm (n-m)\gamma }{2},\\
\cos\alpha^{ee} =& (1-|t_n|^2) \cos\left\{ (n+m)\gamma \right\} +|t_n|^2\cos\varphi, \label{aee}
\end{align}
where $0\leq \alpha^{ee}\leq \pi$.
The Josephson current $J_{ee}$ are also expressed in terms of the energy 
of ABS
\begin{align}
J_{ee}=&J_0 \frac{\Delta_{ }}{\Delta_0} \frac{1}{2T_N} \int_{-\pi/2}^{\pi/2} d\gamma
\frac{\cos(\gamma)|t_n|^2\sin\varphi}{2\sin\alpha^{ee}}\nonumber\\
\times&\left[
\sin(2 X^{ee}_{+}) A(\epsilon^{ee}_+) + \sin(2 X^{ee}_{-}) A(\epsilon^{ee}_-) 
\right].\label{jee}
\end{align}
Next we show the results for a junction where the two superconductors belong to 
the spin-triplet chiral states. 
The two TKNN numbers ($n$ in the left superconductors and $m$ in the right one) 
are odd integer numbers.
We find the energy of ABS as
$E^{\textrm{ABS}}=\epsilon^{oo}_{\pm}$ with,
\begin{align}
\epsilon^{oo}_\pm =&\mp \Delta\textrm{sgn}[ \cos(X^{oo}_\pm) ] \sin(X^{oo}_\pm),\label{oo}\\
X^{oo}_\pm=& \frac{\alpha^{oo}\pm (n-m)\gamma}{2},\\
\cos\alpha^{oo}=&(1-|t_n|^2) \cos\left\{ (n+m)\gamma\right\} - |t_n|^2\cos\varphi.
\end{align}
The Josephson current is calculated as
\begin{align}
J_{oo}=&J_0 \frac{\Delta_{ }}{\Delta_0} \frac{1}{2T_N} \int_{-\pi/2}^{\pi/2} d\gamma
\frac{\cos(\gamma)|t_n|^2\sin\varphi}{2\sin\alpha^{oo}}\nonumber\\
\times&\left[
\sin(2 X^{oo}_{+}) A(\epsilon^{oo}_+) + \sin(2 X^{oo}_{-}) A(\epsilon^{oo}_-)
\right].\label{joo}
\end{align}
Finally we show the results for a junction where the spin-singlet chiral superconductor occupies
the left hand side of the junction and the spin-triplet chiral superconductor 
occupies the right hand side.
The TKNN numbers in the left superconductor $n$ is even integers and
that in the right one $m$ is odd integer numbers.
We find the energy of ABS is $E^{\textrm{ABS}}=\epsilon^{eo}_{\pm}$ with
\begin{align}
\epsilon^{eo}_\pm =&\mp \Delta\textrm{sgn}[ \cos(X^{eo}_\pm) ] \sin(X^{eo}_\pm),\label{eo}\\
X^{eo}_\pm=& \frac{\alpha^{eo}_\nu\pm\{ (n-m)\gamma -\pi/2\} }{2},\\
\cos\alpha^{eo}_\nu=&(1-|t_n|^2) \sin\left\{ (n+m)\gamma\right\} 
+|t_n|^2 \nu \sin\varphi.
\end{align}
The Josephson current becomes 
\begin{align}
J_{eo}=&J_0 \frac{\Delta_{ }}{\Delta_0}\sum_{\nu=\pm1} \frac{1}{2T_N} \int_{-\pi/2}^{\pi/2} d\gamma
\frac{\cos(\gamma)|t_n|^2\nu \cos\varphi}{4\sin\alpha^{eo}_\nu}\nonumber\\
\times&\left[
\sin(2 X^{eo}_{+}) A(\epsilon^{eo}_+) + \sin(2 X^{eo}_{-}) A(\epsilon^{eo}_-)
\right].\label{jeo}
\end{align}
where $\nu =\pm 1$ indicates two Nambu space. 

%
\begin{figure}[th]
\begin{center}
\includegraphics[width=8cm]{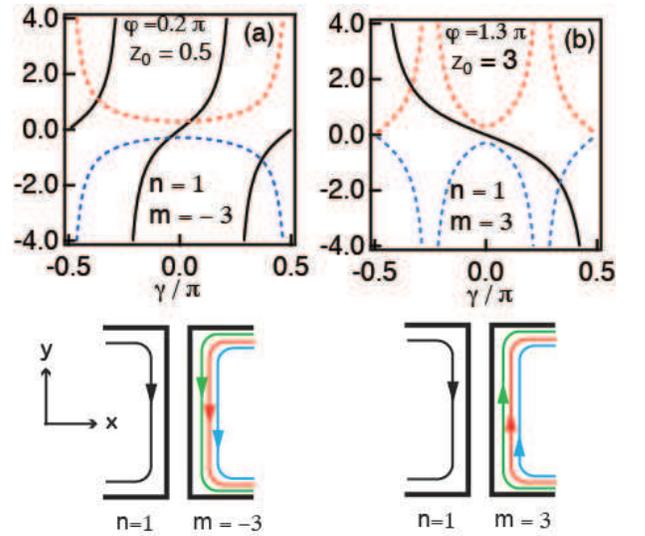}
\end{center}
\caption{(color online). 
The left- and the right-hand side of Eq.~(\ref{ns}) are plotted with a solid line
and two broken lines, respectively. We choose $n=1$, $m=-3$,
$z_0=0.5$, and $\varphi = 0.2\pi$ in (a). The parameters in (b) are 
$n=1$, $m=3$, $z_0=3$, and $\varphi = 1.3\pi$.
}
\label{fig5}
\end{figure}

From the expression of the bound state energy, it is easy to count 
the number of ABS at the zero-energy. For example, we consider 
the junction with both $n$ and $m$ being odd integers here.
At $|t_n|=0$, the energy of the energy of the bound state becomes
\begin{align}
\epsilon^{oo}_+=&-\Delta\textrm{sgn}(\cos n\gamma) \sin n\gamma,\\
\epsilon^{oo}_-=&\Delta\textrm{sgn}(\cos m\gamma) \sin m\gamma,
\end{align}
because of $\alpha^{oo}=(n+m)\gamma$.
They represents the dispersion of the surface bound states when 
the two superconductors are separated from each other. 
It is easy to confirmed that the number of the zero-energy surface states 
of $\epsilon^{oo}_+$ is $|n|$ and that of $\epsilon^{oo}_-$ is $|m|$, 
which is a results of the bulk-boundary 
correspondence for an isolated superconductor.
At finite $|t_n|$,
the solutions of $\epsilon_{oo}=0$ in Eq.~(\ref{oo}) requires
the relation
\begin{align}
\tan\left\{ \frac{(n-m)\gamma}{2} \right\}= \pm \tan\left(\frac{\alpha_{oo}}{2}\right).
\label{ns}
\end{align}
The left hand side of Eq.~(\ref{ns}) goes positive infinity $|n-m|/2$ times and goes 
negative infinity $|n-m|/2$ times within the interval of $-\pi/2 \leq \gamma \leq \pi/2$. 
Since $0\leq\alpha_{oo}\leq\pi$, $\tan\left(\frac{\alpha_{oo}}{2}\right)$ 
on the right hand side remains positive value for all $\gamma$. Thus the number of the 
solutions of Eq.~(\ref{ns}) is $|n-m|$ which corresponds to the difference in the 
TKNN numbers between the two superconductors. 
In Fig.~\ref{fig5}, we plot the left hand 
side of Eq.~(\ref{ns}) with a solid line as a function of $\gamma$ and 
the right hand side with two broken lines. In Fig.~\ref{fig5}(a), we choose 
$n=1$, $m=-3$, $z_0=0.5$, and $\varphi=0.2\pi$. 
The solid line and two broken lines 
cross four times, which means four zero-energy ABS's (i.e., $4=|1-(-3)|$). 
The number of the solutions independent of junction parameters 
such as $z_0$ and $\varphi$. 
In the lower panel in Fig.~\ref{fig5}(a), we illustrate the schematic picture of 
the chiral edge states at $|t_n|=0$. The arrows indicate the direction of chiral current.
In Fig.~\ref{fig5}(b), we choose $n=1$, $m=3$, $z_0=3$, and $\varphi=1.3\pi$.
The results show that there are only two solutions (i.e., $2=|1-3|$). 
In the lower panel, there are four chiral edges at $|t_n|=0$. 
The chiral current in the left superconductor 
flows the opposite direction to that in the right superconductor in this case.
For finite $|t_n|$, two chiral edge current opposite directions are cancel each 
other out and two ABS remain at the zero-energy.
Therefore we confirmed that the number of the ABS at the zero-energy 
is $|n-m|$ for each Nambu space when the TKNN number of two superconductors are $n$ and $m$.
The similar argument can be applied to Eqs.~(\ref{ee}) and (\ref{eo}).
The presence of $|n-m|$ zeros has been suggested by the index theorem~\cite{volovik2}.
Here we count the number of ZES by solving Eq.~(\ref{eq_abs}) explicitly.

The Josephsson current in Eqs.(\ref{jee}) and ({\ref{joo}) have a common 
expression near $T_c$,
\begin{align}
J=&J_0 \frac{\Delta_{ }}{\Delta_0} \frac{\Delta}{T} \sin\varphi 
\frac{1}{4T_N} I_{(n-m)}(z_0),\\
I_{n-m}(z_0)=& \int_{-\pi/2}^{\pi/2} d\gamma
\cos(\gamma)|t_n|^2 \cos\left\{ (n-m)\gamma\right\}
\end{align}
where $n$ and $m$ are both even integer numbers or both odd ones.
The junction is stable at $\pi$-state in the case of $I_{n-m}<0$.
At $z_0=0$, the integral for highly transparent junctions becomes
\begin{align}
I_{n-m}(z_0=0)=\frac{2}{1-(n-m)^2}(-1)^{\frac{n-m}{2}}.
\end{align}
On the other hand in the tunneling limit $z_0\gg 1$, we find
\begin{align}
I_{n-m}(z_0\gg 1)= \frac{4}{z_0^2} \frac{(-1)^{\frac{n-m}{2}}}{ \left\{(n-m)^2-1\right\}
\left\{(n-m)^2-9\right\}}.
\end{align}
In Table.~\ref{table2}, we indicate the stable state of the Josephson junctions
near $T_c$.
At $n-m=4$, $\pi$-state is stable in the highly transparent limit and 
$0$-state is stable in the tunneling limit. Therefore the junction undergoes 
the transition from $\pi$-state to $0$-state when we decrease the transparency 
of the junction. Such junction can be realized with chiral-$d$ wave 
superconductor.
\begin{table}[th]
\caption{\label{table2}%
The stable states of the Josephson junction near $T_c$ are indicated by '0' or '$\pi$'.
Here $n$ and $m$ are both even integer numbers or both odd integer numbers.
As a consequence, $n-m$ becomes an even integer.
}
\begin{ruledtabular}
\begin{tabular}{lcccccc}
$n-m$   & 0 & 2 & 4 & 6 &8 & 10 \\
\colrule
$z_0=0$ & 0 & 0 & $\pi$ & 0 &$\pi$ & 0 \\
$z_0 \gg 1$ & 0 & 0 & 0 & $\pi$ & 0 & $\pi$ \\
\end{tabular}
\end{ruledtabular}
\end{table}

In contrast to Sec.~III, the ABS's at the zero-energy do not affect the Josephson 
current at low temperature. At $T=0$, Eq.~(\ref{jee}) becomes
\begin{align}
J_{ee}=&J_0  \frac{\sin\varphi}{2T_N} \int_{-\pi/2}^{\pi/2} d\gamma
\frac{\cos(\gamma)|t_n|^2\sin\varphi}{2\sin\alpha^{ee}}\nonumber\\
\times&\left[
\frac{\sin(2 X^{ee}_{+})}{|\cos(X^{ee}_+)|}+ \frac{\sin(2 X^{ee}_{-})}{|\cos(X^{ee}_-|)}
\right].\label{jee2}
\end{align}
Mathematically speaking, the zeros in the $|\cos(X^{ee}_\pm)|$ is removed. 
Instead of the zeros of $E^{\text{ABS}}$, minima of $\sin\alpha^{ee}$ in the 
denominator determine the amplitude of Josephson current at low temperature. 
The Josephson current shows the logarithmic dependence of 
temperature at intermediate temperature region between $T=0$ and $T=T_c$.~\cite{barash,ya02}. 
This argument can be applied also to the Josephson current in Eq.~(\ref{joo}).

When $n$ is an even integer and $m$ is an odd integer, 
the Josephsson current in Eqs.(\ref{jeo}) near $T_c$ becomes
\begin{align}
J_{eo}=&-\sin(2\varphi)J_0 \frac{\Delta_{ }}{\Delta_0} \left(\frac{\Delta}{T}\right)^3 \nonumber\\
&\times  
\frac{1}{2T_N} \int_{-\pi/2}^{\pi/2} d\gamma
\frac{\cos(\gamma)|t_n|^4 \cos\left\{ 2(n-m)\gamma\right\}}{96}.
\end{align}
where $n$ and $m$ are an even and an odd integer number, respectively.
The coefficient proportional to $-\sin(2\varphi)$ is positive 
in the limit of both
$z_0=0$ and $z_0\gg 1$.

\begin{figure}[th]
\begin{center}
\includegraphics[width=8cm]{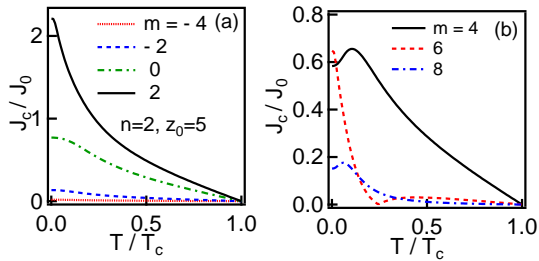}
\end{center}
\caption{(color online). 
The Josephson critical current is plotted as a function of temperature 
for $z_0=5$. We fix TKNN number at $n=2$ in the left superconductor. 
The TKNN number in the right superconductor is $m$
}
\label{fig6}
\end{figure}

In Fig.~\ref{fig6}, we show the maximum value of the Josephson current $J_c$
for $n=2$ and $z_0=5$, where $m$ is the TKNN number on the right superconductor.
The two superconductors belong to spin-singlet symmetry.
The results for $m=-4$, -2, 0, and 2 in Fig.~\ref{fig6}(a) monotonically 
increases with decreasing
temperature. 
On the other hand, the results for $m=4$, 6, and 8 in Fig.~\ref{fig6}(b) 
show the nonmonotonic temperature dependence. 
In particular, the results for $m=6$ indicate the transition
from $0$ state to $\pi$ state with decrease of temperature around $T=0.25T_c$. 
We also note that the $\pi$ state is stable for $m=-4$ and 8 all temperature 
range. The nonmonotonic behavior of $J_c$ indicate the instability 
of the junction between 0 and $\pi$ states. For $m=2$, the Josephson current 
takes its maximum at $\varphi=0.55 \pi$ at $T=0$. In another cases, CPR is 
almost sinusoidal. 
The amplitude of $J_c$ 
becomes smaller for larger $|n-m|$ because the integrand of Eq.~(\ref{jee2})
oscillates more frequently as a function of $\gamma$ for larger $|n-m|$.  

In Fig.~\ref{fig7}, we show the maximum value of the Josephson current $J_c$
for $n=1$ and $z_0=5$, where $m$ is the TKNN number on the right superconductor.
The results for $m=-3$, -1, 1, and 3 in (a) monotonically increase with decreasing
temperature. However, the results for $m=7$ show the nonmonotonic dependence on temperature
because of the 0-$\pi$ transition around $T=0.2T_c$. 
Consistently with the Table.~II, the 0 state is stable at $m\neq 7$ and $z_0\gg 1$
all temperature range. 

\begin{figure}[th]
\begin{center}
\includegraphics[width=8cm]{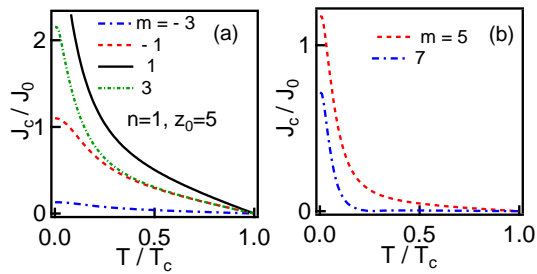}
\end{center}
\caption{(color online). 
The Josephson critical current is plotted as a function of temperature 
for $z_0=5$. 
We fix TKNN number at $n=1$ in the left superconductor. 
The TKNN number in the right superconductor is $m$}
\label{fig7}
\end{figure}

\section{conclusion}
We have studied the properties of the Andreev bound states (ABS) appearing at the junction 
interface between two unconventional superconductors. 
We consider superconductors characterized by two types of topological 
numbers:
the Sato number in Sec.~III and the TKNN number in Sec.~IV.
We confirmed that the number of the ABS's at the zero-energy
is identical to the difference of topological numbers in the two superconductors.
We also discuss effects of the ABS's at the zero-energy on the Josephson current.
The zero-energy ABS's directly contribute to the low-temperature anomaly 
of the Josephson current when the two superconductors characterized by the Sato number.
On the other hand, ABS's at the zero-energy do not affect the Josephson current
 when the two superconductors characterized by the TKNN number.
In the latter case, we also found the 0-$\pi$ transition 
as a function of the transparency of the junction and temperature.

\section{acknowledgement}
This work was supported by KAKENHI(No. 22540355) and 
the "Topological Quantum Phenomena" (No. 22103002) Grant-in Aid for Scientific Research on Innovative Areas from the Ministry of Education, 
Culture, Sports, Science and Technology (MEXT) of Japan.
Y. A. is grateful to Y. Tanaka and M. Sato for useful discussion. 

\end{document}